\documentclass[aps,prl,twocolumn,groupedaddress,superscriptaddress,showpacs,msmath,amssymb]{revtex4}

\usepackage{graphicx}
\usepackage{dcolumn}
\usepackage{bm}
\usepackage[usenames]{color}
\usepackage{epstopdf}
\usepackage[normalem]{ulem}

\begin{document}

\title{
Unoccupied topological surface state in Bi$_{2}$Te$_{2}$Se
}

\author{Munisa Nurmamat}
\affiliation{%
Graduate School of Science, Hiroshima University, 1-3-1 Kagamiyama, Higashi-Hiroshima 739-8526, Japan\\
}

\author{E. E. Krasovskii}
\affiliation{%
Departamento de F\'isica de Materiales UPV/EHU, CFM-MPC UPV/EHU, 20080 San Sebasti\'an/Donostia, Basque Country, Spain\\
}
\affiliation{%
Donostia International Physics Center (DIPC), 20018 San Sebasti$\acute{a}$n/Donostia, Basque Country, Spain\\
}
\affiliation{%
IKERBASQUE, Basque Foundation for Science, 48011 Bilbao, Spain\\
}

\author{K. Kuroda}
\affiliation{%
Graduate School of Science, Hiroshima University, 1-3-1 Kagamiyama, Higashi-Hiroshima 739-8526, Japan\\
}

\author{M. Ye}
\affiliation{%
Shanghai Institute of Microsystem and Information Technology, Chinese Academy of Sciences, Shanghai 200050, China\\
}

\author{K. Miyamoto}
\affiliation{%
Hiroshima Synchrotron Radiation Center, Hiroshima University, 2-313 Kagamiyama, Higashi-Hiroshima 739-0046, Japan\\
}

\author{M. Nakatake}
\affiliation{%
Hiroshima Synchrotron Radiation Center, Hiroshima University, 2-313 Kagamiyama, Higashi-Hiroshima 739-0046, Japan\\
}

\author{T. Okuda}
\affiliation{%
Hiroshima Synchrotron Radiation Center, Hiroshima University, 2-313 Kagamiyama, Higashi-Hiroshima 739-0046, Japan\\
}

\author{H. Namatame}
\affiliation{%
Hiroshima Synchrotron Radiation Center, Hiroshima University, 2-313 Kagamiyama, Higashi-Hiroshima 739-0046, Japan\\
}

\author{H. Namatame}
\affiliation{%
Hiroshima Synchrotron Radiation Center, Hiroshima University, 2-313 Kagamiyama, Higashi-Hiroshima 739-0046, Japan\\
}

\author{M. Taniguchi}
\affiliation{%
Graduate School of Science, Hiroshima University, 1-3-1 Kagamiyama, Higashi-Hiroshima 739-8526, Japan\\
}
\affiliation{%
Hiroshima Synchrotron Radiation Center, Hiroshima University, 2-313 Kagamiyama, Higashi-Hiroshima 739-0046, Japan\\
}

\author{E. V. Chulkov}
\affiliation{%
Departamento de F\'isica de Materiales UPV/EHU, CFM-MPC UPV/EHU, 20080 San Sebasti\'an/Donostia, Basque Country, Spain\\
}
\affiliation{%
Donostia International Physics Center (DIPC), 20018 San Sebasti$\acute{a}$n/Donostia, Basque Country, Spain\\
}

\author{K. A. Kokh}
\affiliation{%
V.S. Sobolev Institute of Geology and Mineralogy, Siberian Branch, Russian Academy of Sciences, Koptyuga pr. 3, Novosibirsk, 630090 Russia\\
}
\affiliation{%
A.V. Rzhanov Institute of Semiconductor Physics, Siberian Branch, Russian Academy of Sciences, 
pr. Akademika Lavrent$^{\prime}$eva 13, Novosibirsk, 630090 Russia\\
}
\affiliation{%
Novosibirsk State University, ul. Pirogova 2, Novosibirsk, 630090 Russia\\
}

\author{O. E. Tereshchenko}
\affiliation{%
A.V. Rzhanov Institute of Semiconductor Physics, Siberian Branch, Russian Academy of Sciences, 
pr. Akademika Lavrent$^{\prime}$eva 13, Novosibirsk, 630090 Russia\\
}
\affiliation{%
Novosibirsk State University, ul. Pirogova 2, Novosibirsk, 630090 Russia\\
}
\affiliation{%
Tomsk State University 634050, Tomsk, Russian Federation\\
}

\author{A. Kimura}
\email{akiok@hiroshima-u.ac.jp}
\affiliation{%
Graduate School of Science, Hiroshima University, 1-3-1 Kagamiyama, Higashi-Hiroshima 739-8526, Japan\\
}

\date{\today}

\begin{abstract}
 Bias voltage dependent scattering of the topological surface state is
studied by scanning tunneling microscopy/spectroscopy for a clean
surface of the topological insulator Bi$_2$Te$_2$Se. A strong warping
of constant energy contours in the unoccupied part of the spectrum 
is found to lead to a spin-selective scattering. The topological 
surface state persists to higher energies in the unoccupied 
range far beyond the Dirac point, where it coexists with the bulk conduction
band. This finding sheds light on the spin and charge dynamics over the 
wide energy range and opens a way to designing opto-spintronic devices.
\end{abstract}


\maketitle

\begin{figure} [t]
\includegraphics{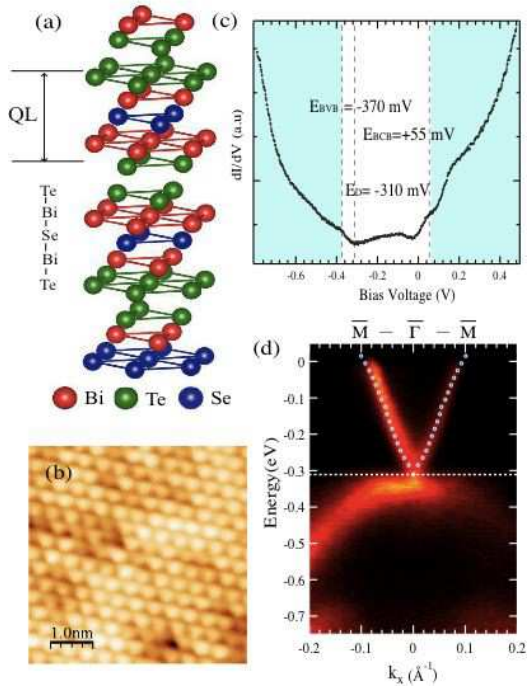}
\caption{ (Color online) (a) Crystal structure showing three quintuple
  units of Bi$_2$Te$_2$Se.  (b) The atomic-resolution image
  (5~nm$\times 5$~nm) on the surface of Bi$_2$Te$_2$Se obtained with
  bias voltage $-50$~mV.  (c) Averaged STS spectrum indicating the
  approximate position for Dirac point (E$_{\rm DP}$), bulk valence
  band (E$_{\rm BVB}$) and bulk conduction band (E$_{\rm BCB}$),
  respectively. (d) Energy-momentum 
  ARPES intensity distribution map along the $\bar{\Gamma}\bar{M}$
  direction for the photon energy of 30~eV.}
\end{figure}

Three-dimensional topological insulators (3D TIs) represent a 
recently discovered state of matter, whose hallmark is the 
surface state in the absolute bulk energy gap,
which has a spin non-degenerate Dirac-cone energy dispersion 
and helical spin texture~\cite{Hasan10,Moore10,Fu07,Hasan11,Qi11,Okuda13}. 
The topological surface state (TSS) is protected by time-reversal symmetry 
and is robust against nonmagnetic perturbations.

Of all the 3D TIs, the most extensively studied is Bi$_2$Se$_3$ 
owing to its large energy gap and the single TSS ~\cite{Xia09,Kuroda10}. However, in spite 
of significant efforts to realize the surface isolated transport in 
Bi$_2$Se$_3$, the progress has been hampered by a too small surface 
contribution to the total conductance compared to the uncontrolled 
bulk contribution from the carrier doping due to the Se 
vacancies~\cite{Butch10,Eto10}.

Here we focus on Bi$_2$Te$_2$Se, which has been theoretically predicted 
to be a 3D TI~\cite{Lin-Lin Wang11} and 
confirmed by angle-resolved photoemission spectroscopy
(ARPES)~\cite{Arakane12, Neupane12, Niesner12}. A highly spin polarized
TSS in Bi$_2$Te$_2$Se has been observed in recent spin-resolved ARPES
(SARPES) measurements~\cite{Miyamoto12}. Bi$_2$Te$_2$Se has an ordered
tetradymite structure, derived from Bi$_2$Te$_3$ by replacing 
the central Te layer with a Se layer. Because here the Se atoms are confined in the central layer 
the formation of Se vacancies and the antisite defects between 
Bi and Te atoms is expected to be less probable~\cite{Ren10},
which would suppress the bulk conductivity. Indeed, this compound has been found bulk insulating, and 
surface-derived quantum oscillations have been observed in 
a magnetotransport experiment~\cite{Ren10}. This makes 
Bi$_2$Te$_2$Se very promising for spintronic 
applications. 

Interband optical excitation of topological surface states by pulsed
laser light is expected to generate longer-lived spin-polarized
carriers at the surface~\cite{Hosur11, Sobota12}. To
understand the photoexcited spin and charge dynamics, knowledge of
unoccupied topological surface state far above the Dirac point and the
unoccupied bulk continuum is crucial. Note that photoelectron
spectroscopy, with which most of the studies on topological insulators
have been performed, cannot access unoccupied states or provide a
direct information on the in-plane electron scattering. Thus, there
has been a dearth of measurements on the unoccupied electronic states
of 3D TIs, and the present study is motivated by the necessity of
getting the information about the unoccupied spectrum.

Scanning tunneling microscopy (STM)/ spectroscopy (STS) has been
widely used to study the surfaces of 3D TIs as the most surface
sensitive technique providing direct information on the electronic structure of topological 
surface states and their scattering properties. One can unveil 
the spin structure of surface states through the presence or absence 
of standing waves both for occupied and unoccupied states. 
Fourier transformed images of the observed standing waves give 
bias-dependent scattering vectors in momentum space. 
For an isotropic TSS, the backscattering is strongly suppressed, 
while a spin-conserving scattering is allowed, and it has actually 
been observed in several topological insulators with warped constant 
energy contours (CECs)~\cite{Fu09,Zhang09,Roushan09,Alpichshev10,
Kim11,Beidenkopf11,Alpichshev12,Ye12}. A non-spin-conserving 
scattering can also occur if the time reversal symmetry is broken, 
i.e., in the presence of magnetic impurities~\cite{Okada11}.

In this Letter, a quasiparticle interference pattern due to surface
scattering is revealed on Bi$_2$Te$_2$Se with a low-temperature STM
experiment over a wide sample bias energy range. A strong warping of
CECs of the TSS explains the anisotropic spin-conserving scattering, 
which persists into the unoccupied state region, where the surface 
state coexists with bulk conduction band.

\begin{figure*}[t]
\includegraphics{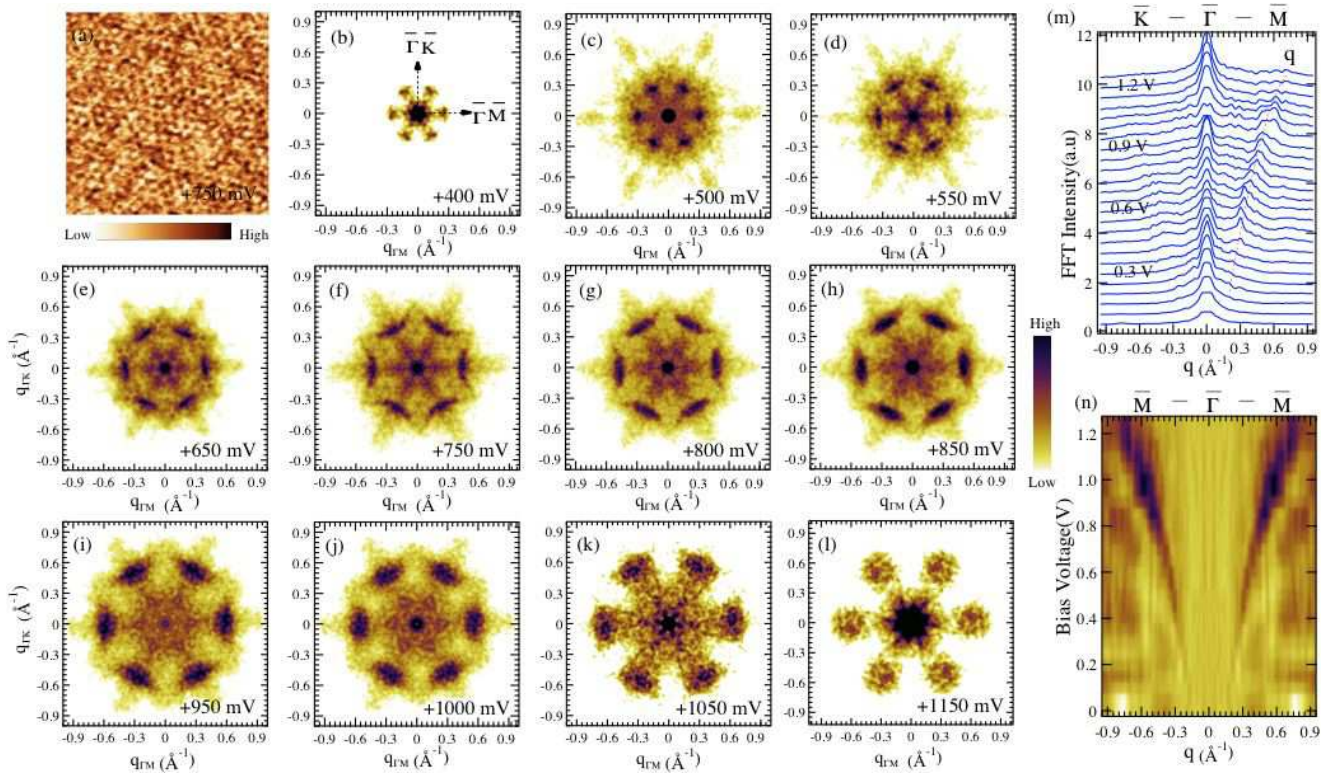}
\caption{ (Color online) (a) Differential conductance map ($dI/dV$
  map) of Bi$_2$Te$_2$Se surface at 4.5 K with sample bias voltage of
  $750$~mV in the 45~nm$\times 45$~nm area (set point
  $I=0.15$~nA). (b)-(l) Fast Fourier transformed images for several
  bias voltages showing six strong spots
  along the $\bar{\Gamma}\bar{M}$ direction. The FFT patterns are
  hexagonally symmetrized.  (m) FFT intensity profiles from $+0.05$~V
  to $+1.25$~V with a step of $+0.05$~V display the bias
  voltage dependent scattering along 
  $\bar{\Gamma}\bar{M}$ (positive $q$) and
  $\bar{\Gamma}\bar{K}$ (negative $q$) directions. (n) The
  ratio of the FFT intensity profiles along $\bar{\Gamma}\bar{M}$ and
  $\bar{\Gamma}\bar{K}$.}
\end{figure*}

Our experiments were performed using an LT-STM (Omicron
NanoTechnology GmbH) operated at 4.5~K in ultrahigh vacuum.
Bi$_2$Te$_2$Se crystal was grown by modified vertical Bridgman 
method as described elsewhere~\cite{Kokh05}. The STM images were 
obtained in a constant-current mode, and the differential conductance 
$dI/dV$ maps were measured simultaneously with recording the 
STM images using a standard lock-in technique. ARPES experiments were
conducted using the synchrotron radiation at BL-7 equipped with a
hemispherical photoelectron analyzer (VG-SCIENTA SES2002) of Hiroshima
Synchrotron Radiation Center (HSRC). Samples were cleaved in
ultra-high vacuum {\it in situ} at room temperature for the STM
and at 10-20 K for ARPES measurement.
 
Similar to Bi$_2$Se$_3$ and Bi$_2$Te$_3$, Bi$_2$Te$_2$Se forms a
rhombohedral crystal structure with the space group $D_{3d}^{5}$
($R\bar3m)$, with the basis quintuple layer (QL) unit of
Te-Bi-Se-Bi-Te, as depicted in Fig.~1(a). Inside the QL the bonds are
predominantly ionic-covalent, and adjacent QLs are bound by van der
Waals forces. Figure~1(b) is the atomic-resolution image
($-50$~mV, $0.12$~nA) of the Bi$_2$Te$_2$Se surface area of
5~nm$\times 5$~nm. The scanning tunneling spectrum gives a measure of
the local density of states near the Fermi energy as shown in
Fig.~1(c). The resulting STS data were averaged over 10 spectra to
improve statistics. The dashed lines show approximate energy locations
of the top of the bulk valence band (BVB), Dirac point (DP) and the
bottom of bulk conduction band (BCB) around the $\bar{\Gamma}$ point.
Figure~1(d) depicts the surface state energy dispersion of
Bi$_2$Te$_2$Se measured by ARPES at the photon energy of $h\nu=30$~eV
[open circles indicate the band dispersion by our {\it ab initio}
calculation (see Fig.~3) shifted downward by 0.24~eV to match the
measured Dirac point position]. The DP energies from ARPES and STS
spectra are equal~\cite{Jia12}.

Figure~2(a) shows the differential conductance ($dI/dV$) map
at a bias voltage of $V_{s}=+750$~mV. It exhibits a standing wave
spreading anisotropically around point defects. (All the spectroscopic
maps at the bias voltages from $+50$ to $+1250$~mV were obtained for
the same surface without changing any other experimental parameters.)
In order to get the momentum space information and obtain the
scattering wave vectors, we have performed Fast Fourier Transformation
(FFT) of the $dI/dV$ maps, see Figs.~2(b)--2(l). These scattering
images provide information on bias-dependent quasiparticle
interference. For bias voltages below $+300$~mV the interference
effect around the point defects is weak, and the FFT image shows a
circular pattern with small $\mathbf{q}$ vectors, which mainly come
from the statistical noise. At $V_{s}=+400$~mV, flower shaped patterns
emerge [Fig.~2(b)] with six broad petals along $\bar{\Gamma}\bar{M}$.
Note that the pattern becomes sharp and intensive at bias voltages
between $+550$ and $+850$~mV. Starting with $V_{s}=+950$~mV, with
increasing the bias voltage the spots get gradually broader.

The evolution of the scattering vectors with $V_{s}$ is visualized 
by the FFT power profiles in Figs.~2(m) and 2(n). In 
Fig.~2(m), in the $\bar{\Gamma}\bar{M}$ direction (rightwards) 
the scattering vectors become larger as $V_{s}$ increases, while there 
is practically no scattering along $\bar{\Gamma}\bar{K}$ (leftwards). 
In Fig.~2(n) we show the ratio of the intensity profiles along
$\bar{\Gamma}\bar{M}$ and $\bar{\Gamma}\bar{K}$. The intensities ratio 
damps the background and makes the scattering more clear: we distinctly 
see the dispersion of $q$ with the bias voltage.

\begin{figure*}
\includegraphics{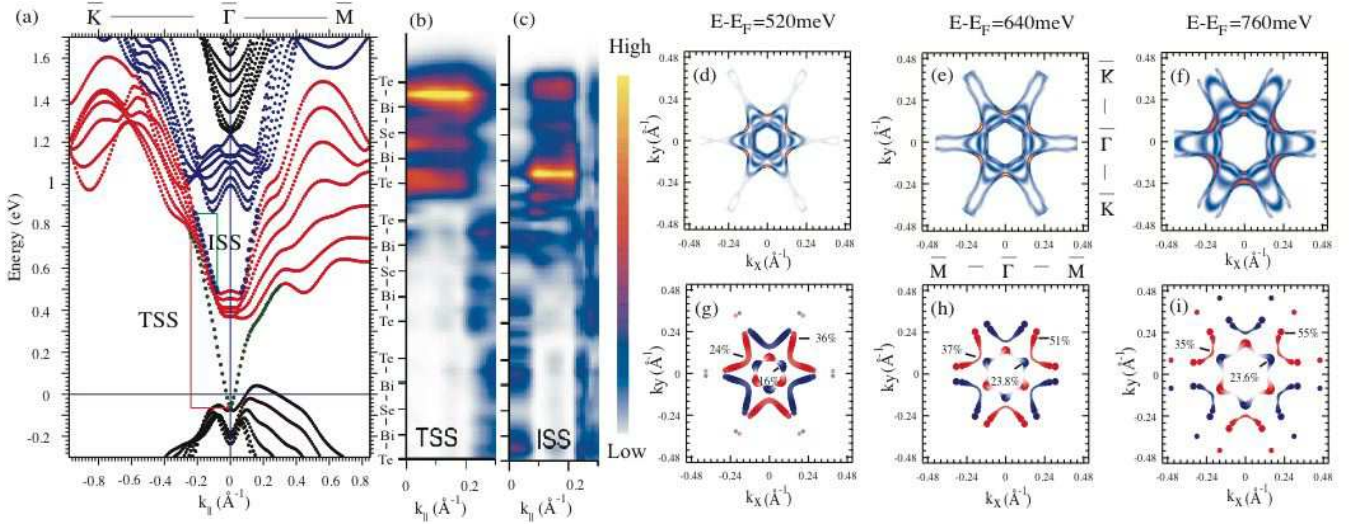}
\caption{ (Color online) (a) First principles electronic structure of
  a 7 formula units slab of Bi$_{2}$Te$_{2}$Se. (b)-(c) The
  depth-momentum distribution of the charge density for topological
  surface state (TSS) and inner surface state (ISS). In view of the
  finite thickness of the slab shown is the sum of the densities of
  the degenerate pair of the surface states located on the opposite
  surfaces of the slab.  (d-f) Spatially resolved Fermi surfaces. The
  color scale shows the constant energy cuts of the surface spectral
  function $ N(z,k_{\parallel})$. (g-i) Distribution of the spin
  polarization perpendicular to the surface, red and blue circles
  denote positive and negative spin polarization and their sizes
  represent the magnitudes of spin polarization. The spin polarization
  values are shown at some specific points. }
\end{figure*}

In order to elucidate the origin of scattering pattern and the effect
of the helical spin texture of the TSS, we have performed a
first-principles calculation of the electronic structure of a 7
formula units slab of Bi$_{2}$Te$_{2}$Se~\cite{method}. Figure~3(a)
shows the band structure along $\bar{\Gamma}\bar{K}$ (leftwards) and
$\bar{\Gamma}\bar{M}$ (rightwards). The magenta arrows show the energy
and momentum ranges of the TSS, and the green arrows indicate the
range of the inner surface state (ISS), which splits off from the
top of the conduction band.  Here, the DP is localized 0.065~eV below
the calculated Fermi energy, i.e., the experimental energy scale is
shifted by 0.24~eV relative to the theoretical scale. 
Figures~3(b) and 3(c) show the depth-momentum distribution 
(in the $\bar{\Gamma}\bar{K}$ direction) of the charge density 
$\rho(z,k_{\parallel})$ for the upper-cone TSS [Fig.~3(b)] and 
for the ISS [Fig.~3(c)]. The upper-cone surface state exists up to $k_{\parallel}=0.22$~\AA$^{-1}$, 
and the ISS between 0.08 and 0.2~\AA$^{-1}$.
 
Figures~3(d)--3(f) show calculated momentum distributions of 
the spatially-resolved spectral density $N(E,{\mathbf k_\parallel})$
at three constant energies $E$. The function is defined as a 
sum over all (discrete) states $\lambda$ with energy $E$ and Bloch 
vector $\mathbf k_\parallel$ weighted with the probability 
$Q_{\lambda{\mathbf k_\parallel}}$ of finding the electron in this 
state in the surface region: 
$N(E,{\mathbf k_\parallel})= \sum_{\lambda}Q_{\lambda
{\mathbf k_\parallel}}\delta(E_{\lambda {\mathbf k_\parallel}}-E)$.
(For the sake of presentation, the $\delta$ function is replaced by 
a Gaussian of 0.05~eV full width at half maximum.) The integral
$Q_{\lambda \mathbf k_\parallel}= \int\! |\psi_{\lambda {\mathbf
 k_\parallel}}(\mathbf r)|^2d{\mathbf r}$ over the surface region
comprises two outermost atomic layers and vacuum.
 
The angular distribution of the spin polarization perpendicular to 
the surface for the two surface states, TSS and ISS, is shown in 
Figs.~3(g)--3(i). Here the net spin density is integrated over a 
half of the slab, and the net spin is normalized to the electron 
charge in the integration region.

The TSS is somewhat stronger localized than the ISS [cf.
Figs.~3(b) and 3(c)], and it exhibits a higher out-of-plane spin
polarization. It is most interesting that the magnitude of the
out-of-plane spin polarization of the TSS may be as large as 55\%.

The bias-dependent quasiparticle scattering is characterized by
scattering vectors that connect the $\mathbf{k}$ vectors of the
initial and final scattering states at the CEC. Figure~4(a) shows
schematic CECs of the TSS. Three characteristic scattering vectors
denoted as $\mathbf{q}_{1}$, $\mathbf{q}_{2}$, and $\mathbf{q}_{3}$
explain the features in Fig.~4(b). The most 
intense croissant-shaped features can only be explained 
by $\mathbf{q}_{2}$ and $\mathbf{q}_{2}^{\prime}$,
which connect two flat segments of the contour as shown in Fig.~4(a).
Other scattering features along the $\bar{\Gamma}\bar{M}$ direction
characterized by $\mathbf{q}_{1}$ and $\mathbf{q}_{3}$, can also be
explained as due to the warping of TSS. The scattering 
originating from ISS can be excluded because its CEC has no 
parallel fragments to cause a large joint density of states, and its 
convex shape does not lead to the croissant-shaped structures. 

To clarify the relation between experimental and theoretical results,
the intensity maxima of the FFT power profiles in the $\bar{\Gamma}\bar{M}$ direction in Fig.~2(m) are compared with 
the $q$ values extracted from the slab calculation, see Fig.~4(c). By 
shifting the calculated points upward by 0.1~eV we were able to reproduce 
all the experimentally observed scattering features. (A discrepancy of 
the same order between the two photon photoemission measurements of the 
unoccupied Dirac cone and calculations was reported in Ref.~[12].)

\begin{figure}
\includegraphics{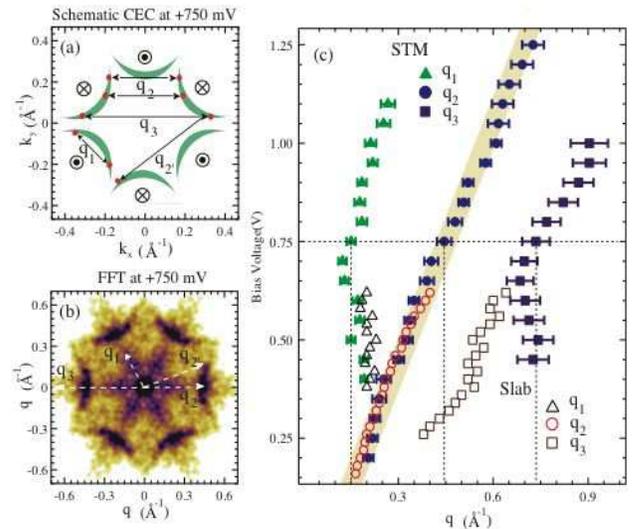}
\caption{ (Color online)(a) Schematic CEC with possible scattering
  vectors and (b) experimental FFT image at a bias voltage of
  $+750$~mV. The $\bar{\Gamma}\bar{M}$ direction is along the
  $x$-axis. (c) Dispersion of three scattering vectors from STM (filled
  symbols) and slab calculation (open symbols).}
\end{figure}


The presence of the FFT features in the $\bar{\Gamma}\bar{M}$
direction and their absence in the $\bar{\Gamma}\bar{K}$ direction
tells us that the scattering is strongly spin selective. This
scattering scenario holds for the whole energy interval from $+300$ to
$+1000$~mV above the Fermi energy, and no significant surface to bulk
scattering is observed, in contrast to  Bi$_2$Se$_3$, for which
a bulk-related scattering has been reported~\cite{Kim11}. 
This indicates that a coupling of the TSS with the bulk continuum states
is negligible even in the unoccupied region, which energetically
overlaps with the bulk conduction band.
 
In conclusion, our scanning tunneling microscopy/spectroscopy
experiment and the first-principles calculation of Bi$_{2}$Te$_{2}$Se
reveal a scattering pattern that originates from the strongly warped constant energy contours of the
topological surface state with substantial out-of-plane spin
polarization. The topological surface state is thus found to survive
up to energies far above the Dirac
point.  This finding provides a deeper understanding of optically excited spin and charge dynamics
at the surface of topological insulators.


STM and ARPES measurements were performed with the approval 
of the Proposal Assessing Committee of HSRC (Proposal No.11-B-40, No.10-A-32).
This work was financially supported by
KAKENHI (Grant No. 20340092, 23340105), Grant-in-Aid for Scientific
Research (B) of JSPS and by RFBR, research project No. 13-02-92105 a. 
The authors acknowledge partial support from
the Spanish Ministerio de Ciencia e Innovaci\'on (Grant No. FIS2010-19609-C02-02).


\begin{thebibliography}{50}

\bibitem{Hasan10} M.Z.~Hasan and C.L.~Kane, Rev. Mod. Phys. {\bf 82}, 3045 (2010).

\bibitem{Moore10} J.E.~Moore, Nature {\bf 464}, 194 (2010).

\bibitem{Fu07} L.~Fu, C.L.~Kane, and E.J.~Mele, Phys. Rev. Lett. {\bf 98}, 106803 (2007).

\bibitem{Qi11} X.L.~Qi and S.C.~Zhang, Rev. Mod. Phys. {\bf 83}, 1057 (2011).

\bibitem{Hasan11}  M.Z.~Hasan and J.E.~Moore, Ann. Review. Condensed Matter Physics.{\bf 2}, 55-78 (2011).

\bibitem{Okuda13} T.~Okuda and A.~Kimura, J. Phys. Soc. Jpn {\bf 82}, 021002 (2013).

\bibitem{Xia09} Y.~Xia, D.~Qian, D.~Hsieh, L.~Wray, A.~Pal, H.~Lin, A.~Bansil, D.~Grauer, Y. S.~Hor, R. J.~Cava, and M. Z.~Hasan,
                Nature Phys. 5, 398 (2009).

\bibitem{Kuroda10} K.~Kuroda, M.~Arita, K.~Miyamoto, M.~Ye, J.~Jiang, A.~Kimura, E.E.~Krasovskii, E.V.~Chulkov, H.~Iwasawa, T.~Okuda,
                     K.~Shimada, Y.~Ueda, H.~Namatame, and M.~Taniguchi, Phys. Rev Lett {\bf 105}, 076802 (2010).

\bibitem{Butch10} N. P.~Butch, K.~Kirshenbaum, P.~Syers, A. B.~Sushkov, G. S.~Jenkins, H. D.~Drew, and J.~Paglione, Phys. Rev. B {\bf 81}, 241301(R) (2010).

\bibitem{Eto10} K.~Eto, Z.~Ren, A. A.~Taskin, K.~Segawa, and Y.~Ando,  Phys. Rev. B {\bf 81}, 195309 (2010).

\bibitem{Lin-Lin Wang11} L.L.~Wang, and D.D.~Johnson, Phys. Rev. B {\bf 83}, 241309(R) (2011)). 

\bibitem{Arakane12} T.~Arakane, T.~Sato, S.~Souma, K.~Kosaka, K.~Nakayama, M.~Komatsu, T.~Takahashi, Z.~Ren, K.~Segawa, and Y.~Ando, Nature Comms. {\bf 3}, 636 (2012).

\bibitem{Neupane12} M.~Neupane, S.-Y.~Xu, L. A.~Wray, A.~Petersen, R.~Shankar, N.~Alidoust, C.~Liu, A.~Fedorov, H.~Ji, J.M.~Allred, Y.S.~Hor, T.R.~Chang, H.T.
                    ~Jeng, H.~Lin, A.~Bansil, R.J.~Cava, and M.Z.~Hasan, Phys. Rev. B {\bf 85}, 235406 (2012).
                     
\bibitem{Niesner12} D.~Niesner, Th.~Fauster, S.V.~Eremeev, T.V.~Menshchikova, Yu.M.~Koroteev, A.P.~Protogenov, E.V.~Chulkov, O.E.~Tereshchenko, K.A.~Kokh,  
                    O.~Alekperov, A.~Nadjafov, and N.~Mamedov, Phys. Rev. B. {\bf 86}, 205403 (2012).
                    
\bibitem{Miyamoto12} K.~Miyamoto, A.~Kimura, T.~Okuda, H.~Miyahara, K.~Kuroda, H.~Namatame, M.~Taniguchi, S.V.~Eremeev, T.V.~Menshchikova, E.V.~Chulkov, K.A.~    
                     Kokh, and O.E.~Tereshchenko, Phys. Rev. Lett. {\bf 109}, 166802 (2012).

\bibitem{Ren10} Z.~Ren, A.A.~Taskin, S.~Sasaki, K.~Segawa, and Y.~Ando, Phys. Rev. B. {\bf 82}, 241306(R) (2010). 

\bibitem{Hosur11} P.~Hosur, Phys. Rev. B {\bf 83}, 035309 (2011)

\bibitem{Sobota12} J.A.~Sobota, S.~Yang, J.G.~Analytis, Y.L.~Chen, I.R.~Fisher, P.S.~Kirchmann, and Z.X.~Shen, Phys. Rev. Lett. {\bf 108}, 117403 (2012)

\bibitem{Zhang09} T.~Zhang,  P.~Cheng,  X.~Chen,  J.F.~Jia,  X.C.~Ma,  K.~He, L.L.~Wang,  H.J.~Zhang,  X.~Dai,  Z.~Fang,  X.C.~Xie, and Q.K.~Xue, Phys. Rev. 
                  Lett. {\bf 103}, 266803 (2009).
                  
\bibitem{Fu09}   L.~Fu, Phys. Rev. Lett. 103, 266801 (2009).                  

\bibitem{Roushan09} P.~Roushan, J.~Seo, C.V.~Parker, Y.S.~Hor, D.~Hsieh, D.~Qian, A.~Richardella, M.Z.~Hasan, R.J.~Cava and A.~Yazdani, Nature {\bf 460}, 1106   
                    (2009). 

\bibitem{Alpichshev10} Z.~Alpichshev,  J.G.~Analytis,  J.H.~Chu,  I.R.~Fisher,  Y.L.~Chen,  Z.X.~Shen,  A.~Fang, and  A.~Kapitulnik, Phys. Rev. Lett. {\bf  
                       104}, 016401 (2010).

\bibitem{Kim11} S.~Kim, M.~Ye, K.~Kuroda, Y.~Yamada, E.E.~Krasovskii, E.V.~Chulkov, K.~Miyamoto, M.~Nakatake, T.~Okuda, Y.~Ueda, K.~Shimada, H.~Namatame, M.~
                Taniguchi, and A.~Kimura, Phys. Rev. Lett. {\bf 107}, 056803 (2011).

\bibitem{Beidenkopf11} H.~Beidenkopf, P.~Roushan, J.~Seo, L.~Gorman, I.~Drozdov, Y.S.~Hor, R.J.~Cava and A.~Yazdani, Nature {\bf 7}, 939-943 (2011).

\bibitem{Alpichshev12} Z.~Alpichshev,  R.R.~Biswas,  A.V.~Balatsky,  J.G.~Analytis,  J.H.~Chu,  I.R.~Fisher, and A.~Kapitulnik, Phys. Rev. Lett. {\bf 108}, 
                       206402 (2012).

\bibitem{Ye12} M.~Ye, S.V.~Eremeev, K.~Kuroda, E.E.~Krasovskii, E.V.~Chulkov, Y.~Takeda, Y.~Saitoh, K.~Okamoto, S.Y.~Zhu, K.~Miyamoto, M.~Arita, M.~Nakatake, T.~
               Okuda, Y.~Ueda, K.~Shimada, H.~Namatame, M.~Taniguchi, and A.~Kimura, Phys. Rev. B, {\bf 85}, 205317 (2012).
               
\bibitem{Okada11} Y.~Okada, C.~Dhital, W.W.~Zhou, E.D.~Huemiller, H.~Lin, S.~Basak, A.~Bansil, Y.B.~Huang, H.~Ding, Z.~Wang, S.D.~Wilson,  
                  and V.~Madhavan, Phys. Rev. Lett. {\bf 106}, 206805 (2011).               

\bibitem{Kokh05} K.A.~Kokh, B.G.~Nenashev, A.E.~Kokh, and G.Y.~Shvedenkov, J. Crystal Growth {\bf 275}, e2129 (2005).

\bibitem{Jia12} S.~Jia, H.~Beidenkopf, I.~Drozdov, M.K.~Fuccillo, J.~Seo,J.~Xiong, N.P.~Ong, A.~Yazdani, and R.J.~Cava Phys. Rev. B. {\bf 86}, 165119 (2012).

\bibitem{method} Self-consistent calculation within the local density approximation
were performed with the full-potential augmented plane wave method described in 
E.E.~Krasovskii, F.~Starrost, and W.~Schattke, Phys. Rev. B {\bf 59}, 10504 (1999).


\end{thebibliography}
\end{document}